%% file: main.tex
\documentclass[a4paper]{article}
\usepackage{INTERSPEECH2018}
\usepackage{cite,url}

\title{The fifth `CHiME' Speech Separation and Recognition Challenge: Dataset, task and baselines}
\name{Jon Barker$^1$, Shinji Watanabe$^2$, Emmanuel Vincent$^3$, and Jan Trmal$^2$}
\address{
  $^1$University of Sheffield, UK\\
  $^2$Center for Language and Speech Processing, Johns Hopkins University, Baltimore, USA\\
  $^3$Universit\'e de Lorraine, CNRS, Inria, LORIA, F-54000 Nancy, France}
\email{j.p.barker@sheffield.ac.uk, shinjiw@jhu.edu, emmanuel.vincent@inria.fr, jtrmal@gmail.com}

\begin{document}

\maketitle
 
\begin{abstract}
The CHiME challenge series aims to advance robust automatic speech recognition (ASR) technology by promoting research at the interface of speech and language processing, signal processing, and machine learning. This paper introduces the 5th CHiME Challenge, which considers the task of distant multi-microphone conversational ASR in real home environments. Speech material was elicited using a dinner party scenario with efforts taken to capture data that is representative of natural conversational speech and recorded by 6 Kinect microphone arrays and 4 binaural microphone pairs. The challenge features a single-array track and a multiple-array track and, for each track, distinct rankings will be produced for systems focusing on robustness with respect to distant-microphone capture vs.\ systems attempting to address all aspects of the task including conversational language modeling. We discuss the rationale for the challenge and provide a detailed description of the data collection procedure, the task, and the baseline systems for array synchronization, speech enhancement, and conventional and end-to-end ASR.\\
\end{abstract}


\section{Introduction}
\label{sec:intro}
\input{introduction.tex}

\section{Dataset}
\label{sec:data}
\input{dataset.tex}

\section{Task}
\label{sec:task}
\input{task.tex}

\section{Baselines}
\label{sec:baselines}
\input{baselines.tex}

\section{Conclusion}
\label{sec:concl}
The `CHiME' challenge series is aimed at evaluating ASR in real-world conditions. This paper has presented the 5th edition which targets conversational speech in an informal dinner party scenario recorded with multiple microphone arrays. The full dataset and state-of-the-art software baselines have been made publicly available. A set of challenge instructions has been carefully designed to allow meaningful comparison between systems and maximise scientific outcomes. The submitted systems and the results will be announced at the 5th `CHiME' ISCA Workshop.

\section{Acknowledgements}
We would like to thank Google for funding the full data collection and annotation, Microsoft Research for providing Kinects, and Microsoft India for sponsoring the 5th `CHiME' Workshop. E.\ Vincent acknowledges support from the French National Research Agency in the framework of the project VOCADOM ``Robust voice command adapted to the user and to the context for AAL'' (ANR-16-CE33-0006). 

\bibliographystyle{IEEEtran}

\bibliography{mybib}

\end{document}

%% file: introduction.tex
Automatic speech recognition (ASR) performance in difficult reverberant and noisy conditions has improved tremendously in the last decade \cite{virtanen2012book,Li2015,newera,Vincent2018,Makino2018}. This can be attributed to advances in speech processing, audio enhancement, and machine learning, but also to the availability of real speech corpora recorded in cars \cite{Aurora3,CUMove}, quiet indoor environments \cite{TED,MC-WSJ-AV_PASCAL_SSC2_2012_MMA_REVERB_RealData_2}, noisy indoor and outdoor environments \cite{CHiME3,CHiME4}, and challenging broadcast media \cite{ETAPE,MGB}. Among the applications of robust ASR, voice command in domestic environments has attracted much interest recently, due in particular to the release of Amazon Echo, Google Home and other devices targeting home automation and multimedia systems. The CHiME-1 \cite{Barker2013} and CHiME-2 \cite{Vincent2013ASRU12} challenges and corpora have contributed to popularizing research on this topic, together with the DICIT \cite{DICIT_1}, Sweet-Home \cite{SweetHome}, and DIRHA \cite{Ravanelli2015} corpora. These corpora feature single-speaker reverberant and/or noisy speech recorded or simulated in a single home, which precludes the use of modern speech enhancement techniques based on machine learning. The recently released voiceHome corpus \cite{voiceHome} addresses this issue, but the amount of data remains fairly small.

In parallel to research on acoustic robustness, research on conversational speech recognition has also made great progress, as illustrated by the recent announcements of super-human performance \cite{Xiong2017,Saon2017} achieved on the Switchboard telephone conversation task \cite{Switchboard} and by the ASpIRE challenge \cite{harper2015automatic}. Distant-microphone recognition of noisy, overlapping, conversational speech is now widely believed to be the next frontier. Early attempts in this direction can be traced back to the ICSI \cite{ICSI}, CHIL \cite{CHIL}, and AMI \cite{AMI} meeting corpora, the LLSEC \cite{LLSEC} and COSINE \cite{COSINE} face-to-face interaction corpora, and the Sheffield Wargames corpus \cite{SWC}. These corpora were recorded using advanced microphone array prototypes which are not commercially available, and as result could only be installed in a few laboratory rooms. The Santa Barbara Corpus of Spoken American English \cite{SBC} stands out as the only large-scale corpus of naturally occurring spoken interactions between a wide variety of people recorded in real everyday situations including face-to-face or telephone conversations, card games, food preparation, on-the-job talk, story-telling, and more. Unfortunately, it was recorded via a single microphone.

The CHiME-5 Challenge aims to bridge the gap between these attempts by providing the first large-scale corpus of real multi-speaker conversational speech recorded via commercially available multi-microphone hardware in multiple homes. Speech material was elicited using a 4-people dinner party scenario and recorded by 6 distant Kinect microphone arrays and 4 binaural microphone pairs in 20 homes. The challenge features a single-array track and a multiple-array track. Distinct rankings will be produced for systems focusing on acoustic robustness vs.\ systems aiming to address all aspects of the task.

The paper is structured as follows. Sections \ref{sec:data} and \ref{sec:task} describe the data collection procedure and the task to be solved. Section \ref{sec:baselines} presents the baseline systems for array synchronization, speech enhancement, and ASR and the corresponding results. We conclude in Section \ref{sec:concl}.

%% file: dataset.tex
\subsection{The scenario}

The dataset is made up of the recording of twenty separate dinner parties taking place in real homes. Each dinner party has four participants - two acting as hosts and two as guests. The party members are all friends who know each other well and who are instructed to behave naturally.

Efforts have been taken to make the parties as natural as possible. The only constraints are that each party should last a minimum of 2 hours and should be composed of three phases, each corresponding to a different location: i) \textit{kitchen} -- preparing the meal in the kitchen area; ii) \textit{dining} -- eating the meal in the dining area; iii) \textit{living} -- a post-dinner period in a separate living room area.

Participants were allowed to move naturally from one location to another but with the instruction that each phase should last at least 30 minutes. Participants were left free to converse on any topics of their choosing. Some personally identifying material was redacted post-recording as part of the consent process. Background television and commercial music were disallowed in order to avoid capturing copyrighted content.

\subsection{Audio}

Each party has been recorded with a set of six Microsoft Kinect devices. The devices have been strategically placed such that there are always at least two capturing the activity in each location. Each Kinect device has a linear array of 4 sample-synchronised microphones and a camera. The raw microphone signals and video have been recorded. Each Kinect is recorded onto a separate laptop computer. Floor plans were drafted to record the layout of the living space and the approximate location and orientation of each Kinect device.

In addition to the Kinects, to facilitate transcription, each participant wore a set of Soundman OKM II Classic Studio binaural microphones. The audio from these was recorded via a Soundman A3 adapter onto Tascam DR-05 stereo recorders also being worn by the participants.


\subsection{Transcriptions}

The parties have been fully transcribed. For each speaker a reference transcription is constructed in which, for each utterance produced by that speaker, the start and end times and the word sequence are manually obtained by listening to the speaker's binaural recording (the reference signal). For each other recording device, the utterance's start and end time are produced by shifting the reference timings by an amount that compensates for the a synchrony between devices (see Section \ref{sec:sync}).

The transcriptions can also contain the followings tags: $[noise]$ denoting any non-language noise made by the speaker (e.g., coughing, loud chewing, etc.); $[inaudible]$ denoting speech that is not clear enough to be transcribed; $[laughs]$ denoting instances of laughter; $[redacted]$ are parts of the signals that have been zeroed out for privacy reasons.

%% file: task.tex
\label{s:task}
\subsection{Training, development, and evaluation sets}

The 20 parties have been divided into disjoint training, development and evaluation sets as summarised in Table \ref{t:data}. There is no overlap between the speakers in each set.

\begin{table}[tbh]
\caption{Overview of CHiME-5 datasets}
\label{t:data}
\centering
\begin{tabular}{c|c|c|c|c} 
\toprule
Dataset & Parties & Speakers & Hours & Utterances \\ 
\midrule
Train & 16 & 32 & 40:33 & 79,980 \\ 
Dev & 2 & 8 & 4:27 & 7,440 \\ 
Eval & 2 & 8 & 5:12 & 11,028 \\ 
\bottomrule
\end{tabular}
\end{table}

For the development and evaluation data, the transcription file also contains a speaker location and `reference' array for each utterance. The location can be either `kitchen',`dining room', or `living room' and the reference array (the target for speech recognition) is chosen to be one that is situated in the same area.

\subsection{Tracks and ranking}
\label{s:tracks}
The challenge features two tracks:
\begin{itemize}
\item \textit{single-array}: only the reference array can be used to recognise a given evaluation utterance,
\item \textit{multiple-array}: all arrays can be used.
\end{itemize}
For each track, two separate rankings will be produced:
\begin{itemize}
\item \textit{Ranking A} -- systems based on conventional acoustic modeling and using the supplied official language model: the outputs of the acoustic model must remain frame-level tied phonetic (senone) targets and the lexicon and language model must not be modified,
\item \textit{Ranking B} -- all other systems, e.g., including systems based on end-to-end processing or systems whose lexicon and/or language model have been modified.
\end{itemize}
In other words, ranking A focuses on acoustic robustness only, while ranking B addresses all aspects of the task.

\subsection{Instructions}

A set of instructions has been provided that ensure that systems are broadly comparable and that participants respect the application scenario. In particular, systems are allowed to exploit knowledge of the utterance start and end time, the utterance speaker label and the speaker location label. During evaluation participants can use the entire session recording from the reference array (for the single-array track) or from all arrays (for multiple-array track), i.e., one can use the past and future acoustic context surrounding the utterance to be recognised. For training and development, participants are also provided with the binaural microphone signals and the floor plans. Participants are forbidden from manual modification of the data or the annotations (e.g., manual refinement of the utterance timings or transcriptions). 

It is required that all parameters are tuned on the training set or development set. Participants can evaluate different versions of their system but the final submission will be the one that performs best on the development data, and this will be ranked according to its performance on the evaluation data. While, some modifications of the development set are necessarily allowed (e.g. automatic signal enhancement, or refinement of utterance timings), participants have been cautioned against techniques designed to fit the development data to the evaluation data (e.g. by selecting subsets, or systematically varying its pitch or level). These ``biased'' transformations are forbidden.

The challenge has been designed to promote research covering all stages in the recognition pipeline. Hence, participants are free to replace or improve any component of the baseline system, or even to replace the entire baseline with their own systems. However, the architecture of the system will determine whether a participant's result is ranked in category A or category B (see Section \ref{s:tracks}).

Participants will evaluate their own systems and will be asked to return overall WERs for the development and evaluation data, plus WERs broken down by session and location. They will also be asked to submit the corresponding lattices in Kaldi format to allow their scores to be validated, plus a technical description of their system.

%% file: baselines.tex
\subsection{Array synchronization}
\label{sec:sync}
While signals recorded by the same device are sample-synchronous, there is no precise synchronisation between devices. Across devices synchronisation cannot be guaranteed. The signal start times are approximately synchronised post-recording using a synchronisation tone that was played at the beginning of each recording session. However, devices can drift out of synchrony due to small variations in clock speed (clock drift) and due to frame dropping. To correct for this a cross-correlation approach is used to estimate the delay between one of the binaural recorders chosen as the reference and all other devices \cite{Knapp1976}. These delays are estimated at regular 10 second intervals throughout the recording. Using the delay estimates, separate utterance start and end times have been computed for each device and are recorded in the JSON transcription files.

\subsection{Speech enhancement}
\label{sec:se}
CHiME-5 uses a weighted delay-and-sum beamformer (BeamformIt \cite{Anguera2007}) as a default multichannel speech enhancement approach, similar to the CHiME-4 recipe \cite{CHiME4}.
Beamforming is performed by using four microphone signals attached to the reference array.
The reference array information is provided by the organizers through the JSON transcription file.

\subsection{Conventional ASR}
The conventional ASR baseline is distributed through the Kaldi github repository \cite{Povey_ASRU2011_2011}\footnote{\url{https://github.com/kaldi-asr/kaldi/tree/master/egs/chime5/s5}} and is described in brief below.

\subsubsection{Data preparation (stage 0 and 1)}
\label{sec:data_prep_kaldi}
These stages provide Kaldi-format data directories, lexicons, and language models.
We use a CMU dictionary\footnote{\url{http://www.speech.cs.cmu.edu/cgi-bin/cmudict}} as a basic pronunciation dictionary.
However, since the CHiME-5 conversations are spontaneous speech and a number of words are not present in the CMU dictionary, we use grapheme to phoneme conversion based on Phonetisaurus G2P \cite{novak2012wfst}\footnote{\url{https://github.com/AdolfVonKleist/Phonetisaurus}} to provide the pronunciations of these OOV (out-of-vocabulary) words.
The language model is selected automatically, based on perplexity on training data, but at the time of the writing, the selected LM is 3-gram trained by the MaxEnt modeling method as implemented in the SRILM toolkit \cite{junwu, srilm-me, stolcke2002srilm}. The total vocabulary size is 128K augmented by the G2P process mentioned above.

\subsubsection{Enhancement (stage 2)} 
This stage calls BeamformIt based speech enhancement, as introduced in Section \ref{sec:se}.

\subsubsection{Feature extraction and data arrangement (stage 3-6)}

These stages include MFCC-based feature extraction for GMM training, and training data preparation (250k utterances, in \verb|data/train_worn_u100k|)
The training data combines both left and right channels (150k utterances) of the binaural microphone data (\verb|data/train_worn|) and a subset (100k utterances) of all Kinect microphone data (\verb|data/train_u100k|).
Note that we observed some performance improvements when we use larger amounts of training data instead of the above subset. 
However, we have limited the size of the data in the baseline so that experiments can be run without requiring unreasonable computational resources.

\subsubsection{HMM/GMM (stage 7-16)}
Training and recognition are performed with a hidden Markov model (HMM) / Gaussian mixture model (GMM) system. 
The GMM stages include standard triphone-based acoustic model building with various feature transformations including linear discriminant analysis (LDA), maximum likelihood linear transformation (MLLT), and feature space maximum likelihood linear regression (fMLLR) with  speaker adaptive training (SAT).

\subsubsection{Data cleanup (stage 17)}

This stage removes several irregular utterances, which improves the final performance of the system \cite{peddinti2016far}.
Totally 15\% of utterances in the training data are excluded due to this cleaning process, which yields consistent improvement in the following LF-MMI TDNN training.

\subsubsection{LF-MMI TDNN (stage 18)}
\label{sec:lm_mmi}
This is an advanced time-delayed neural network (TDNN) baseline using lattice-free maximum mutual information (LF-MMI) training \cite{Povey2016}. 
This baseline requires much larger computational resources: multiple GPUs for TDNN training (18 hours with 2-4 GPUs), many CPUs for i-vector and lattice generation, and large storage space for data augmentation (speed perturbation).

As a summary, compared with the previous CHiME-4 baseline \cite{CHiME4}, the CHiME-5 baseline introduces: 1) grapheme to phoneme conversion; 2) Data cleaning up; 3) Lattice free MMI training.
With these techniques, we can provide a reasonable ASR baseline for this challenging task.  

\subsection{End-to-end ASR}
CHiME-5 also provides an end-to-end ASR baseline based on ESPnet\footnote{\url{https://github.com/espnet/espnet}}, which uses Chainer \cite{chainer} and PyTorch \cite{pytorch}, as its underlying deep learning engine.

\subsubsection{Data preparation (stage 0)}
This is the same as the Kaldi data directory preparation, as discussed in Section \ref{sec:data_prep_kaldi}.
However, the end-to-end ASR baseline does not require lexicon generation and FST preparation.
This stage also includes beamforming based on the BeamformIt toolkit, as introduced in Section \ref{sec:se}.

\subsubsection{Feature extraction (stage 1)}
This stage use the Kaldi feature extraction to generate Log-Mel-filterbank and pitch features (totally 83 dimensions).
It also provides training data preparation (350k utterances, in \verb|data/train_worn_u200k|), which combines both left and right channels (150k utterances) of the binaural microphone data (\verb|data/train_worn|) and a subset (200k utterances) of all Kinect microphone data (\verb|data/train_u200k|).

\subsubsection{Data conversion for ESPnet (stage 2)}
This stage converts all the information included in the Kaldi data directory (transcriptions, speaker IDs, and input and output lengths) to one JSON file (\verb|data.json|) except for input features.
This stage also creates a character table (45 characters appeared in the transcriptions).

\subsubsection{Language model training (stage 3)}
Character-based LSTM language model is trained by using either a Chainer or PyTorch backend, which is integrated with a decoder network in the following recognition stage.

\subsubsection{End-to-end model training (stage 4)}
A hybrid CTC/attention-based encoder-decoder network \cite{Watanabe2017} is trained by using either the Chainer or PyTorch backend.
The total training time is 12 hours with a single GPU (TitanX) when we use the PyTorch backend, which is less than the computational resources required for the Kaldi LF-MMI TDNN training (18 hours with 2-4 GPUs).

\subsubsection{Recognition (stage 5)}
Speech recognition is performed by combining the LSTM language model and end-to-end ASR model trained by previous stages with multiple CPUs.

\subsection{Baseline results}
Tables \ref{tab:binaural} and \ref{tab:array} provide the word error rates (WERs) of the binaural (oracle) and reference Kinect array (challenge baseline) microphones.
\begin{table}[tbh]
  \caption{WERs for the development set using the binaural microphones (oracle).}
  \label{tab:binaural}
  \centering
  \begin{tabular}{ l | c}
    \toprule
     & Development set \\
   	\midrule
    conventional (GMM) & 72.8 \\
	conventional (LF-MMI TDNN)	 & 47.9 \\
	end-to-end & 67.2 \\
    \bottomrule
  \end{tabular}
\end{table}
\begin{table}[tbh]
  \caption{WERs for the development set using the reference Kinect array with beamforming (challenge baseline).
  }
  \label{tab:array}
  \centering
  \begin{tabular}{ l | c}
    \toprule
     & Development set \\
   	\midrule
    conventional (GMM) & 91.7 \\
	conventional (LF-MMI TDNN)	 & 81.3 \\
	end-to-end & 94.7 \\
    \bottomrule
  \end{tabular}
\end{table}
The WERs of the challenge baseline are quite high due to very challenging environments of CHiME-5 for all of methods\footnote{Note, the current end-to-end ASR baseline performs poorly due to an insufficient amount of training data. 
However, the result of end-to-end ASR was better than that of the Kaldi GMM system when we used the binaural microphones for testing, which shows end-to-end ASR to be a promising direction for this challenging environment.}.
Comparing these tables, there is a significant performance difference between the array and binaural microphone results (e.g., 33.4\% absolutely in LF-MMI TDNN), which indicates that the main difficulty of this challenge comes from the source and microphone distance in addition to the spontaneous and overlapped nature of the speech, which exist in both array and binaural microphone conditions.
So, a major part of the challenge lies in developing speech enhancement techniques that can improve the challenge baseline to the level of the binaural microphone performance. 

Table \ref{tab:room} shows the WER of the LF-MMI TDNN system with the development set for each session and room.
The challenge participants have to submit this form with the evaluation set.
We observe that performance is poorest in the kitchen condition, probably due to the kitchen background noises and greater degree of speaker movement that occurs in this location.
\begin{table}[tbh]
  \caption{WERs of the LF-MMI TDNN system for each session and room conditions.
  The challenge participants have to submit this form scored with the evaluation set.}
  \label{tab:room}
  \centering
  \begin{tabular}{l|cc}
  \toprule

        & \multicolumn{2}{l}{Development set} \\
        & S02              & S09              \\
  \midrule
KITCHEN & 87.3             & 81.6             \\
DINING  & 79.5             & 80.6             \\
LIVING  & 79.0             & 77.6             \\
  \bottomrule
\end{tabular}
\end{table}